\documentclass{ws-procs975x65}

\usepackage{hyperref}
\usepackage{graphicx}

\def\vct#1{{\mathbf{#1}}}
\def\neanl{\nonumber\\}

\def\canmomp{p}

\def\vcanmom{\hat{\vct{\canmomp}}}

\newcommand{\vnunit}{{\vnxa{12}}}
\newcommand{\vmom}[1]{{\vcanmom}_{#1}}
\newcommand{\vnxa}[1]{{\vct{n}}_{#1}}

\newcommand{\scpm}[2]{(#1\,#2)}

\newcommand{\vspin}[1]{\hat{\vct{S}}_{#1}}
\newcommand{\relab}[1]{r_{#1}}
\newcommand{\rel}{\relab{12}}
\newcommand{\ang}{L}
\newcommand{\vang}{\vct{\ang}}

\allowdisplaybreaks

\begin{document}

\title{\uppercase{Recent progress in spin calculations in the post-Newtonian framework and applications}}

\def\Fadr{Theoretisch--Physikalisches Institut, \\
	Friedrich--Schiller--Universit\"at, \\
	Max--Wien--Platz 1, 07743 Jena, Germany, EU}

\def\Sadr{Centro Multidisciplinar de Astrof\'isica (CENTRA), Departamento de F\'isica, \\
	Instituto Superior T\'ecnico (IST), Universidade T\'ecnica de Lisboa, \\ %(UTL)
	Avenida Rovisco Pais 1, 1049-001 Lisboa, Portugal, EU}

\author{\uppercase{Johannes Hartung}}
\address{TPI, Friedrich--Schiller--Universit\"at, Max--Wien--Platz 1, 07743 Jena, Germany
\\johannes.hartung@uni-jena.de}

\author{\uppercase{Jan Steinhoff}}
\address{CENTRA, Instituto Superior T\'ecnico, Avenida Rovisco Pais 1, 1049-001 Lisboa, Portugal}

\author{\uppercase{Gerhard Sch\"afer}}
\address{TPI, Friedrich--Schiller--Universit\"at, Max--Wien--Platz 1, 07743 Jena, Germany}

\begin{abstract}
Recently we derived the next-to-next-to-leading order post-Newtonian Hamiltonians at spin-orbit and spin(1)-spin(2) level for a binary system of compact objects. In this talk the derivation of them will be shortly outlined at an introductory level. We will also discuss some checks of our (complicated and long) results in the first part of the talk. In the second part we will show how to apply our results to the calculation of the last stable circular orbit of such a binary system of black holes or neutron stars.
\end{abstract}

\keywords{Post-Newtonian approximation; Spinning binaries; Equations of motion}

\bodymatter\bigskip

Astrophysical binary systems usually contain spinning components. With the dawn of gravitational
wave astronomy approaching, spin contributions to the equations of motion become very important.
Recent years have seen a lot of progress in this direction, see Ref.~\citen{Hartung:Steinhoff:Schafer:2012}
for a summary of literature in this field. The present work uses an extension of the ADM formalism\cite{Arnowitt:Deser:Misner:1962}
to spinning objects linear in the spins\cite{Steinhoff:Schafer:2009:2,Steinhoff:2011}. The Hamiltonians in question
are the next-to-next-to-leading order (NNLO) spin-orbit (SO) and NNLO spin(1)-spin(2) (SS) ones
\cite{Hartung:Steinhoff:2011:1,Hartung:Steinhoff:2011:2}. Details on the corresponding formal 3PN calculations will be covered in 
a recent manuscript\cite{Hartung:Steinhoff:Schafer:2012,Hartung:2012}.

% overview of calculation method, dim-reg etc.
% 
% \rmk{Vorlage, kuerzen:} The next-to-next-to-leading order post-Newtonian spin-orbit and spin(1)-spin(2) Hamiltonians for 
% binary compact objects in general relativity are derived. The Arnowitt-Deser-Misner canonical formalism and 
% its generalization to spinning compact objects in general relativity are presented and a fully reduced matter-only 
% Hamiltonian is obtained. Several simplifications using integrations by parts are discussed.
% Approximate solutions to the constraints and evolution equations of motion are provided. Technical details of the integration
% procedures are given including an analysis of the short-range behavior of the integrands around the sources.
% The Hamiltonian of a test-spin moving in a stationary Kerr spacetime is obtained by rather simple approach
% and used to check parts of the mentioned results. Kinematical consistency checks by using the global 
% (post-Newtonian approximate) Poincar\'{e} algebra are applied.
% Along the way a self-contained overview for the computation of the 3PN ADM point-mass Hamiltonian is provided, too.

%\section{Results and Checks}
For brevity we provide the Hamiltonians in the center-of-mass frame. The full results are given in the original
publications\cite{Hartung:Steinhoff:2011:1,Hartung:Steinhoff:2011:2,Hartung:Steinhoff:Schafer:2012,Hartung:2012}.
In this frame and in dimensionless quantities\cite{Tessmer:Hartung:Schafer:2010,Tessmer:Hartung:Schafer:2012}
they are given by
\begin{align}
%=======================================================================
H_{\text{COM SO}}^{\text{NNLO}}
&= \frac{1}{4\rel^5} \left[
    21\sqrt{1-4\eta}(\eta+1)\scpm{\vang}{\vct{\Delta}}
    +\frac{1}{2} (-2\eta^2+33\eta+42)\scpm{\vang}{\vct{\Sigma}}
  \right] \neanl&
  + \frac{\eta}{32 \rel^4} \biggl[
    -\sqrt{1-4\eta}\left((256+45\eta)\scpm{\vnunit}{\vmom{}}^2 + (314+39\eta)\vmom{}^2\right)\scpm{\vang}{\vct{\Delta}} \neanl & \quad
    +\left((-256+275\eta)\scpm{\vnunit}{\vmom{}}^2 + (-206+73\eta)\vmom{}^2\right)\scpm{\vang}{\vct{\Sigma}}
  \biggr] \neanl&
  + \frac{\eta}{32 \rel^3} \biggl[
    \sqrt{1-4\eta}\bigl(
	15\scpm{\vnunit}{\vmom{}}^4
	+ 3(9\eta-4) \scpm{\vnunit}{\vmom{}}^2 \vmom{}^2 \neanl & \quad\quad
	+ 2(22\eta-9) (\vmom{}^2)^2
    \bigr)\scpm{\vang}{\vct{\Delta}} 
    -\bigl(
	15(2\eta-1)\scpm{\vnunit}{\vmom{}}^4 \neanl &\quad\quad
	+3 (6\eta^2 - 11 \eta + 4) \scpm{\vnunit}{\vmom{}}^2 \vmom{}^2
	+2 (5\eta^2 - 3\eta + 2) (\vmom{}^2)^2
    \bigr)\scpm{\vang}{\vct{\Sigma}}
  \biggr]\,,\\
% %=======================================================================
H_{\text{COM SS}}^{\text{NNLO}}
&=
\eta\biggl\{
  \frac{1}{4\rel^5} \left[
    (79\eta + 105) \scpm{\vnunit}{\vspin{1}}\scpm{\vnunit}{\vspin{2}}
    -(63 + 19\eta) \scpm{\vspin{1}}{\vspin{2}}
  \right] \neanl &\quad
    +\frac{1}{\rel^4} \biggl[
      -\left(
	\frac{303}{4}\eta \scpm{\vnunit}{\vmom{}}^2 
	+ \left(\frac{125}{4}\eta+9\right) \vmom{}^2
      \right) \scpm{\vnunit}{\vspin{1}}\scpm{\vnunit}{\vspin{2}} \neanl & \quad\quad
      \left(
      	-\left(18+\frac{25}{4}\eta\right) \scpm{\vnunit}{\vmom{}}^2 
	+ \left(9 + \frac{47}{2}\eta\right) \vmom{}^2
      \right) \scpm{\vspin{1}}{\vspin{2}} \neanl & \quad\quad
      - \frac{9}{4} (7\eta+4) \scpm{\vmom{}}{\vspin{1}}\scpm{\vmom{}}{\vspin{2}} \neanl & \quad\quad
      +\left(34\eta+\frac{27}{2}\right)\scpm{\vnunit}{\vmom{}}
	(\scpm{\vmom{}}{\vspin{1}}\scpm{\vnunit}{\vspin{2}} + \scpm{\vnunit}{\vspin{1}}\scpm{\vmom{}}{\vspin{2}}) \neanl & \quad\quad
      +\frac{3}{2}\sqrt{1-4\eta}(\eta+3)\scpm{\vnunit}{\vmom{}}
	(\scpm{\vmom{}}{\vspin{1}}\scpm{\vnunit}{\vspin{2}} - \scpm{\vnunit}{\vspin{1}}\scpm{\vmom{}}{\vspin{2}})
    \biggr] \neanl &\quad
   +\frac{1}{\rel^3} \biggl[
      \frac{1}{8}\biggl(
	105 \eta^2 \scpm{\vnunit}{\vmom{}}^4
	+15 \eta (3\eta - 2) \scpm{\vnunit}{\vmom{}}^2 \vmom{}^2 \neanl &\quad\quad\quad 
	+\frac{3}{2} (10\eta^2 + 13 \eta - 6) (\vmom{}^2)^2
      \biggr)\scpm{\vnunit}{\vspin{1}}\scpm{\vnunit}{\vspin{2}} \neanl &\quad\quad
      +\frac{1}{8}\biggl(
	-3 (8\eta^2 - 37 \eta + 12) \scpm{\vnunit}{\vmom{}}^2 \vmom{}^2 
	+ (7\eta^2 - 23 \eta +9) (\vmom{}^2)^2
      \biggr)\scpm{\vspin{1}}{\vspin{2}} \neanl &\quad\quad
      +\frac{1}{4}\biggl(
	9 \eta^2 \scpm{\vnunit}{\vmom{}}^2 
	+ \frac{1}{2} (4\eta^2 + 25 \eta - 9) \vmom{}^2
      \biggr)\scpm{\vmom{}}{\vspin{1}}\scpm{\vmom{}}{\vspin{2}} \neanl &\quad\quad
      -\frac{3}{8}\biggl(
	+ 15 \eta^2 \scpm{\vnunit}{\vmom{}}^2 
	+ \frac{1}{2} (10\eta^2 + 21 \eta - 9) \vmom{}^2
      \biggr)\scpm{\vnunit}{\vmom{}}\neanl&
	\quad\quad\quad\quad\times(\scpm{\vmom{}}{\vspin{1}}\scpm{\vnunit}{\vspin{2}} + \scpm{\vnunit}{\vspin{1}}\scpm{\vmom{}}{\vspin{2}}) \neanl &\quad\quad
      +\frac{9}{16}\sqrt{1-4\eta}(1 - 2\eta)\scpm{\vnunit}{\vmom{}}(\scpm{\vmom{}}{\vspin{1}}\scpm{\vnunit}{\vspin{2}} - \scpm{\vnunit}{\vspin{1}}\scpm{\vmom{}}{\vspin{2}})
   \biggr]
\biggr\}\,.
%=======================================================================
\end{align}
There $\vct{\Delta} = \vspin{1}-\vspin{2}$ and $\vct{\Sigma} = \vspin{1}+\vspin{2}$ the differences and sums of 
the spin vectors, $\vang$ is the orbital angular momentum $\vang = \rel \vnunit \times \vmom{}$ and $\eta = m_1 m_2/(m_1 + m_2)^2$ is
the symmetric mass ratio. The spins $\vspin{a}$, relative position $\vct{r}_{12}$ and linear 
momentum $\vmom{}$ variables fulfill the usual canonical Poisson brackets, where $\rel = |\vct{r}_{12}|$ and $\vnunit = \vct{r}_{12}/\rel$.

Certain tests of these results were performed. Most important, kinematical consistency checks by using the global 
(post-Newtonian approximate) Poincar\'{e} algebra are applied. This requires the appropriate center-of-mass 
vectors\cite{Hartung:Steinhoff:2011:1, Hartung:Steinhoff:2011:2, Hartung:Steinhoff:Schafer:2012}.
Furthermore, we checked our Mathematica code by re-deriving the 3PN ADM point-mass Hamiltonian including the
corresponding ultraviolet analysis in dimensional regularization.
Also the Hamiltonian of a test-spin moving in a stationary Kerr spacetime is 
obtained by rather simple approach and used to check parts of the mentioned results.

%\section{Application: Corrections to the last stable circular orbit (LSO)}

\begin{figure}
\begin{center}
\includegraphics{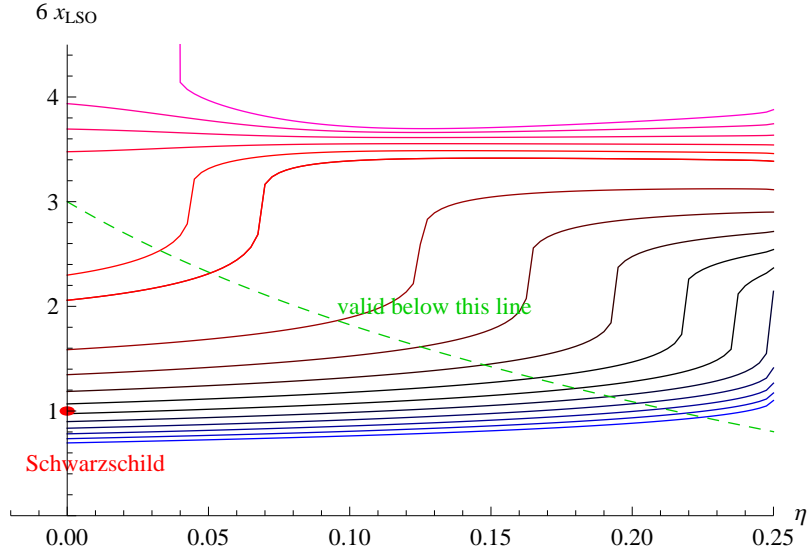}
\end{center}
 \caption{Last stable circular orbit for $S=0.1$ plotted for different symmetric mass ratios $\eta$ and Kerr spins $a$. 
 (From lower to upper lines $a=-1.0$ to $a=0.84$. The two uppermost plots contain the cases $a=0.8$ and $a=0.84$ respectively. The difference in Kerr spin
 between all other plots is $\Delta a = 0.2$.) The ellipse on the vertical axis denotes the last stable circular orbit of a testmass orbiting a Schwarzschild black hole.\label{fig:lso}}
\end{figure}

As an application the last stable circular orbit was determined via an ansatz modified by $\eta$-dependent coefficients, which
were matched to the post-Newtonian expansion derived from the mentioned Hamiltonians\cite{Tessmer:Hartung:Schafer:2012},
see Fig. \ref{fig:lso}. Notice that the approximation brakes down above the dashed line.

\emph{Acknowledgments ---}
This work is supported by the Deutsche Forschungsgemeinschaft (DFG) through projects
GRK 1523 ``Quanten- und Gravitationsfelder,'' STE 2017/1-1,  and SFB/TR7.
Furthermore we thank M.~Tessmer who is coauthor of two of the most important works
for this talk for useful and enlighting discussions.
% We gratefully acknowledge P.~Jaranowski for sharing insight in 3PN point-mass calculations, for useful discussions
% on integration procedures, for providing several testintegrals, and for useful discussions on
% calculating the spin-precession frequency. 
% In particular we thank T.~Damour for very useful comments and hints regarding the calculation
% of $\dim$-dimensional short-range behavior of certain integrals. 
% We additionally thank M. Levi for pleasant collaboration on the comparison of our result 
% with the potential obtained within the effective field theory approach.
% We also thank S.~Hergt and M.~Tessmer for many
% useful discussions about the Poincar\'{e} algebra and orbital parameterization issues, respectively.
% We further gratefully acknowledge useful discussions 
% with D.~Brizuela on {\scshape xPert} and doing perturbation theory in 
% arbitrary dimensions,
% with M.~Q.~Huber on three-body integral related Appell $F_4$ functions,
% with H.~Witek on numerical relativity and literature on some special topics, and 
% with T.~J.~Rothe on implementation issues in {\scshape Mathematica} and for giving
% us some hints to mathematical theorems. 

% \bibliographystyle{utphys}
% \bibliographystyle{ws-procs975x65}
% \bibliography{../references}

\providecommand{\href}[2]{#2}\begingroup\raggedright\endgroup

% generated by bibtex, taken from main.bbl. Added arXiv numbers.
% \begin{thebibliography}{1}

% \bibitem{Hartung:Steinhoff:Schafer:2012}
% J.~Hartung, J.~Steinhoff and G.~Sch{\"a}fer, \texttt{arXiv:1302.6723 [gr-qc]}.

% \bibitem{Arnowitt:Deser:Misner:1962}
% R.~L. Arnowitt, S.~Deser and C.~W. Misner, The dynamics of general relativity,
%   in {\em Gravitation: An Introduction to Current Research\/},  ed. L.~Witten
%   (John Wiley, New York, 1962) pp. 227--265.

% \bibitem{Steinhoff:Schafer:2009:2}
% J.~Steinhoff and G.~Sch{\"a}fer, {\em Europhys. Lett.} {\bf 87}, p. 50004
%   (2009).

% \bibitem{Steinhoff:2011}
% J.~Steinhoff, {\em Ann. Phys. (Berlin)} {\bf 523}, 296 (2011).

% \bibitem{Hartung:Steinhoff:2011:1}
% J.~Hartung and J.~Steinhoff, {\em Ann. Phys. (Berlin)} {\bf 523}, 783 (2011).

% \bibitem{Hartung:Steinhoff:2011:2}
% J.~Hartung and J.~Steinhoff, {\em Ann. Phys. (Berlin)} {\bf 523}, 919 (2011).

% \bibitem{Hartung:2012}
% J.~Hartung, {B}in{\"a}rsysteme kompakter {O}bjekte in hoher postnewtonscher
%   {N}{\"a}herung in den {E}igendrehimpuls-{W}echselwirkungen, PhD thesis,
%   {Friedrich-Schiller-Universit{\"a}t Jena}, 2012.

% \bibitem{Tessmer:Hartung:Schafer:2010}
% M.~Tessmer, J.~Hartung and G.~Sch{\"a}fer, {\em Class. Quant. Grav.} {\bf 27},
%   p. 165005 (2010).

% \bibitem{Tessmer:Hartung:Schafer:2012}
% M.~Tessmer, J.~Hartung and G.~Sch{\"a}fer, {\em Class. Quant. Grav.} {\bf 30},
%   p. 015007 (2013).

% \end{thebibliography}

\end{document}